\documentclass[aps,prl,twocolumn,superscriptaddress, amsmath,amssymb]{revtex4-2}

\usepackage{soul}
\usepackage{diagbox}
\usepackage{ragged2e}
\usepackage{array}
\usepackage{booktabs}
\usepackage{graphicx}
\usepackage{dcolumn}
\usepackage{bm}
\usepackage[colorlinks, linkcolor=black, anchorcolor=black, citecolor=black]{hyperref}

\begin{document}

\preprint{APS/123-QED}

\title{Universal Spectral Mirage Gaps in Superconductors with Time-Reversal-Symmetric Spin-Orbit Coupling}

\author{Xusheng Wang}
\email{wang-xs23@mails.tsinghua.edu.cn}
\affiliation{%
State Key Laboratory of Low-Dimensional Quantum Physics, Department of Physics, Tsinghua University, Beijing 100084, China}
\author{Gaomin Tang}
\email{gmtang@gscaep.ac.cn}
\affiliation{
Graduate School of China Academy of Engineering Physics, Beijing 100193, China}
\author{Shuai-Hua Ji}
\affiliation{%
State Key Laboratory of Low-Dimensional Quantum Physics, Department of Physics, Tsinghua University, Beijing 100084, China}
\affiliation{%
Frontier Science Center for Quantum Information, Beijing 100084, China}

\date{\today}

\begin{abstract}

Magnetic-field-induced spectral mirage gaps, regarded as evidence of finite-energy pairing correlations, have so far been mainly studied in superconductors with Ising spin-orbit coupling (SOC). Here, we show that superconductors with any time-reversal-symmetric SOC can induce mirage gaps near the SOC energy scale when the applied magnetic field has a component perpendicular to the SOC texture, whereas the parallel component produces Zeeman-split spectral features near the superconducting gap. We demonstrate this general principle in superconductors with Rashba and Rashba-Ising SOC. These universal field-dependent signatures establish superconducting spectroscopy as a powerful probe of SOC textures and strengths.


\end{abstract}

\maketitle
Magnetic fields interact with superconductivity primarily through orbital and Zeeman effects \cite{fulde1973high}. In two-dimensional superconductors under an in-plane field, orbital depairing is strongly suppressed, providing an ideal platform for investigating the interplay between superconductivity and Pauli-paramagnetic effects. Two central manifestations of this interplay are the magnetic-field--temperature phase diagram and the quasiparticle spectrum. For conventional superconductors, the phase diagram is characterized by a zero-temperature critical field known as the Pauli limit \cite{clogston1962upper,chandrasekhar1962note,sarma1963influence,maki1966effect}, while the spectrum exhibits Zeeman-split superconducting coherence peaks  \cite{meservey1970magnetic,meservey1975tunneling,gubbels2013imbalanced}, as extensively studied both theoretically and experimentally \cite{meservey1994spin}.

Superconductors with spin-orbit interactions, which are ubiquitous in solids \cite{amundsen2024colloquium}, exhibit different magnetic responses. In disordered superconductors, strong spin-orbit scattering can substantially enhance the zero-temperature critical field \cite{klemm1975theory,tedrow1982critical} and suppress the spectral Zeeman splitting of superconducting coherence peaks \cite{alexander1985theory,swartz2018superconducting}. In clean superconductors, intrinsic spin-orbit coupling (SOC) can not only enhance the Pauli limit \cite{gor2001superconducting,barzykin2002inhomogeneous,olde2021tunable,ilic2017enhancement,mockli2020ising}, but also give rise to topological superconducting excitations \cite{tewari2011topologically,zhou2016ising,schirmer2024topological}. Notably, SOC in nonmagnetic superconductors, including Ising- and Rashba-type SOC, generally preserves time-reversal symmetry (TRS) \cite{ghosh2021recent}. Ising superconductivity \cite{zhu2011giant,yuan2016ising,liu2017unconventional}, which has been widely observed in transition-metal dichalcogenide monolayers \cite{lu2015evidence,saito2016superconductivity,xi2016ising,sohn2018unusual,yi2022crossover,de2018tuning,cho2022nodal,li2021recent,wan2023orbital,ji2024continuous,volavka2026ising}, has attracted considerable interest because of its strong enhancement of the in-plane critical field \cite{ilic2017enhancement,mockli2020ising,liu2020microscopic}. However, transport measurements often cannot provide decisive low-temperature evidence, since the corresponding critical fields can be experimentally inaccessible. Theoretical studies have proposed the spectral mirage gaps, accessible by low-temperature tunneling measurements, as direct signatures of Ising superconductivity \cite{tang2021magnetic,ilic2023spectral}. In realistic devices, interfacial hybridization and electrostatic gating \cite{lu2015evidence,saito2016superconductivity} can introduce additional TRS SOC components, such as Rashba SOC \cite{wang2015strong,masseroni2024spin}. However, spectral consequences of superconductors with general TRS SOC remain poorly understood.


In this Letter, we theoretically investigate the spectra of spin-singlet $s$-wave superconductors with TRS SOC under a Zeeman-type magnetic field. We demonstrate that any TRS SOC can produce spectral mirage gaps near its characteristic energy scale when the applied magnetic field has a component perpendicular to the SOC texture, whereas the parallel component gives rise to Zeeman-split spectral features. The momentum dependence of the SOC modifies the shape of the mirage gaps but does not eliminate it, as illustrated by representative examples with Rashba and Ising SOC. Our results establish spectral mirage gaps as a universal probe of SOC textures and strengths in superconductors.

\begin{figure*}[htbp]
\centering
\includegraphics[width=\linewidth]{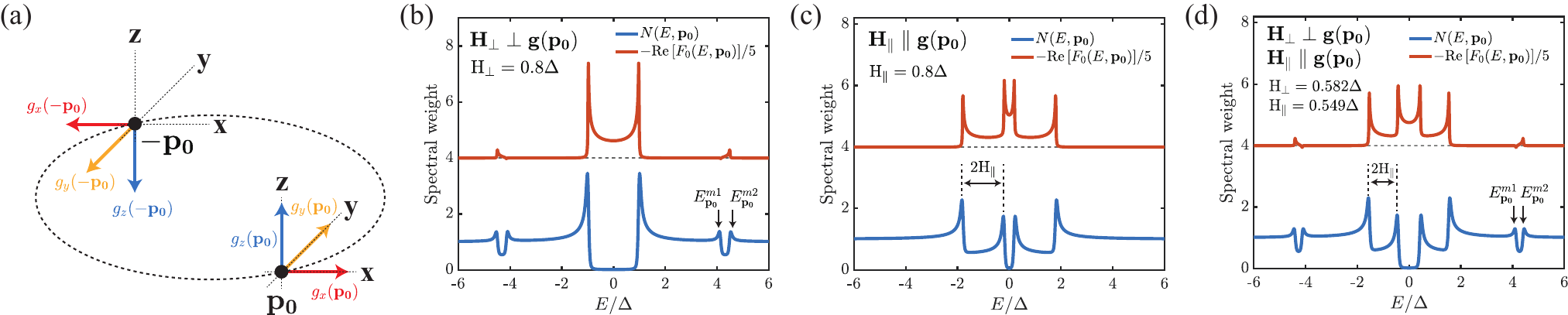}
\caption{
(a) Schematic spin texture of a general TRS SOC at $\pm \mathbf{p_0}$ near the Fermi level. The SOC fields at opposite momenta are related by time-reversal symmetry $\mathbf{g}(\mathbf{p})=-\mathbf{g}(\mathbf{-p})$.
(b-d) DOS and singlet pairing correlation $-\mathrm{Re}(F_0)/5$ for (b) $\mathbf{H}\perp \mathbf{g}(\mathbf{p_0})$, (c) $\mathbf{H}\parallel \mathbf{g}(\mathbf{p_0})$ and (d) $\mathbf{H}$ with both perpendicular $\mathbf{H_{\perp}}$ and parallel $\mathbf{H_{\parallel}}$ components relative to $\mathbf{g}(\mathbf{p_0})$. We use $\mathbf{g}(\mathbf{p_0})=(2\Delta,2\Delta,3\Delta)$, $|\mathbf{H}|=0.8\Delta$, and $\eta=0.01\Delta$. A small thermal broadening $k_BT=0.02\Delta$ is included. The $-\mathrm{Re}(F_0)$ curves are vertically offset by 4 and scaled by a factor of $1/5$ for clarity.
}
\label{general_illustration}
\end{figure*}

For a spin-singlet $s$-wave superconductor with a general TRS SOC [Fig.~\ref{general_illustration}(a)], the normal-state Hamiltonian under an in-plane magnetic field reads \cite{lu2015evidence}
\begin{equation}\label{normal_state}
    \hat{H}_N(\mathbf{p})=\xi_\mathbf{p}\sigma_0+\mathbf{g}(\mathbf{p})\cdot\boldsymbol{\sigma} - \mathbf{H}\cdot\boldsymbol{\sigma},
\end{equation}

\noindent where $\xi_\mathbf{p}$ is the kinetic energy relative to the Fermi level. The vector $\mathbf{g}(\mathbf{p})=-\mathbf{g}(\mathbf{-p})$ describes the momentum-dependent spin texture of a general TRS SOC \cite{Frigeri2004}. The $\boldsymbol{\sigma}=(\sigma_x,\sigma_y,\sigma_z)$ and $\sigma_0$ denote the Pauli matrices and identity matrix in spin space, respectively. The Zeeman energy is $\mathbf{H}=g_L\mu_B \mathbf{B}/2$, where $g_L$, $\mu_B$, and $\mathbf{B}=(B_x,B_y,0)$ denote the Land\'e $g$ factor, Bohr magneton, and in-plane magnetic field, respectively. 

In the Nambu basis $(\hat{c}_{\mathbf{p},\uparrow}, \hat{c}_{\mathbf{p},\downarrow}, \hat{c}_{-\mathbf{p},\uparrow}^\dagger, \hat{c}_{-\mathbf{p},\downarrow}^\dagger)$,
the Bogoliubov-de Gennes Hamiltonian is \cite{harms2025collapse,patil2023spectral,wang2026first}:
\begin{equation}\label{BdG_eq}
    \hat{H}_{\mathrm{BdG}}(\mathbf{p})=    
    \begin{pmatrix}
    \hat{H}_N(\mathbf{p}) & i\Delta \sigma_y \\
    -i\Delta \sigma_y & -\hat{H}_N(-\mathbf{p})
    \end{pmatrix}.
\end{equation}
The momentum-resolved spectral properties and pairing correlations are characterized by the normal Green's function $G(E,\mathbf{p})$ and anomalous Green's function $F(E,\mathbf{p})$, respectively. They are given by
\begin{equation}
    \begin{pmatrix}
    G(E,\mathbf{p}) & F(E,\mathbf{p})  \\
    \Bar{F}(E,\mathbf{p}) & \Bar{G}(E,\mathbf{p})
    \end{pmatrix}=\left[E+i\eta - \hat{H}_{\mathrm{BdG}}(\mathbf{p})\right]^{-1}.
\end{equation}
Here, $\eta$ is a small phenomenological broadening parameter \cite{balatsky2006impurity,dynes1978direct}. The anomalous Green's function is decomposed into
\begin{equation}
    F(E,\mathbf{p})=(F_0\sigma_0 + F_x\sigma_x + F_y\sigma_y + F_z\sigma_z)i\sigma_y,
\end{equation}
\noindent where $F_0$ represents the spin-singlet pairing correlation, while $F_x$, $F_y$, and $F_z$ represent the spin-triplet components. The momentum-resolved density of states (DOS) is computed as $\rho(E,\mathbf{p})=-\mathrm{Im}\{\mathrm{Tr} [G(E,\mathbf{p})]\}/\pi$, which is thermally broadened into $N(E,\mathbf{p})$ \cite{supp}. The overall pairing correlations and DOS are obtained by integrating over momentum space.

Here, we focus on the DOS of a superconductor with general TRS SOC at a fixed momentum $\mathbf{p}_0$. At $\mathbf{p_0}$, the SOC spin texture is $\mathbf{g}(\mathbf{p_0})=(g_x(\mathbf{p_0}), g_y(\mathbf{p_0}), g_z(\mathbf{p_0}))$. We first consider a magnetic field perpendicular to the spin texture. In this case, diagonalizing Eq.~(\ref{BdG_eq}) gives the quasiparticle dispersion

{\small
\begin{equation}\label{H_perp}
    E_{\mathbf{p_0}}(\mathbf{H}_{\perp})=\pm\sqrt{\mathrm{H}_{\mathrm{eff}}^2+\Delta^2+\xi_{\mathbf{p_0}}^2\pm 2\sqrt{\mathrm{H_{\perp}^2}\Delta^2+\mathrm{H}_{\mathrm{eff}}^2\xi_{\mathbf{p_0}}^2}},
\end{equation}
}

\noindent where $\mathrm{H}_{\mathrm{eff}}=\sqrt{|\mathbf{g}(\mathbf{p_0})|^2+\mathrm{H_{\perp}^2}}$ is the effective magnetic field. Equation ~(\ref{H_perp}) closely resembles the quasiparticle dispersion of an Ising superconductor \cite{tang2021magnetic}, indicating a mirage-gap center at $E_{\mathbf{p0}}^{mc}=(E_{\mathbf{p0}}^{m1}+E_{\mathbf{p0}}^{m2})/2$, where $E_{\mathbf{p0}}^{m2(m1)}=\sqrt{|\mathbf{g}(\mathbf{p_0})|^2+(\mathrm{H_{\perp}}\pm\Delta)^2}$ denotes the energy of the mirage-gap coherence peak. The presence of the mirage gaps and their pairing correlations are further shown by Fig.~\ref{general_illustration}(b). Since the triplet-pairing correlations are coordinate dependent, only the coordinate-independent singlet-pairing correlation is presented.

For a magnetic field parallel to the spin texture, the quasiparticle dispersion is
{\small
\begin{equation}\label{H_para}
    E_{\mathbf{p_0}}(\mathbf{H}_{\parallel})=\pm \sqrt{\Delta^2+(\xi_{\mathbf{p_0}}+s |\mathbf{g}(\mathbf{p_0})|)^2}+ s\mathrm{H}_{\parallel},\ 
\end{equation}
}

\noindent with $s=\pm 1$. Equation~(\ref{H_para}) indicates Zeeman-split-like superconducting coherence peaks (see Supplemental Material \cite{supp}), similar to those in conventional superconductors \cite{meservey1970magnetic,meservey1975tunneling,gubbels2013imbalanced}. This behavior is supported by the numerical results in Fig.~\ref{general_illustration}(c).

For a magnetic field with both perpendicular and parallel components, the quasiparticle dispersion becomes more complicated. Nevertheless, the spectral response can be understood by decomposing the field into $\mathbf{H}_{\perp}$ and $\mathbf{H}_{\parallel}$: the perpendicular component produces the mirage gaps described by Eq.~(\ref{H_perp}), whereas the parallel component gives rise to the Zeeman-split-like behavior described by Eq.~(\ref{H_para}). This picture is confirmed by the numerical results in Fig.~\ref{general_illustration}(d). Since the mirage-gap position encodes $|\mathbf{g}(\mathbf{p}_0)|$ and the Zeeman-splitting strength reflects the relative orientation between the magnetic field and the SOC direction, these spectral signatures provide a direct probe of SOC textures and strengths. In addition, the nonzero anomalous Green's function inside the mirage gaps further demonstrates the presence of finite-energy pairing. Notably, the overall DOS requires integration over momentum space, which can modify the shape of the mirage gaps but does not eliminate them. Therefore, mirage gaps can be generally induced by an appropriate magnetic field in superconductors with any TRS SOC.

\begin{figure*}[ht]
\centering
\includegraphics[width=0.95\linewidth]{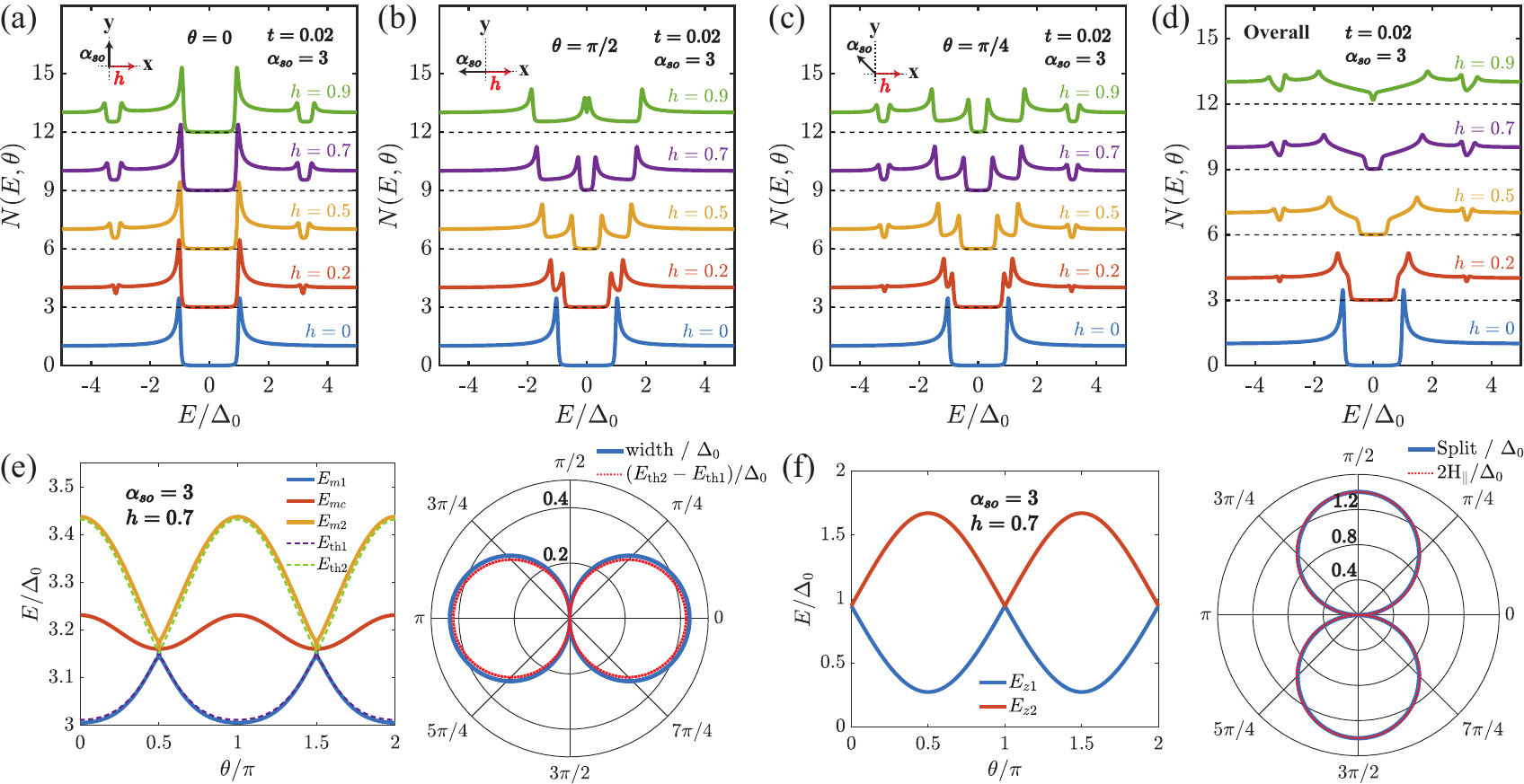}
\caption{
Magnetic-field-dependent spectral features of a Rashba superconductor with $\alpha=3$. Field-dependent momentum-resolved DOS $N(E,\mathbf{p})$ at (a) $\theta=0$, (b) $\theta=\pi/2$, and (c) $\theta=\pi/4$. Black arrows indicate the spin directions pinned by Rashba SOC, and red arrows indicate the magnetic-field direction. Successive DOS curves are vertically offset by 3 for clarity.
(d) Overall field-dependent DOS $N(E)$.
(e) Angular dependence of the mirage gap at $h=0.7$. The associated coherence peaks are denoted by $E_{m1}$ and $E_{m2}$, which are fitted by $E_{\mathrm{th1}}=\sqrt{(\alpha p_F)^2+(\mathrm{H_\perp}-\Delta)^2}/\Delta_0$ and $E_{\mathrm{th2}}=\sqrt{(\alpha p_F)^2+(\mathrm{H_\perp}+\Delta)^2}/\Delta_0$. The mirage-gap center is denoted by $E_{mc}$. The mirage-gap width is given by $E_{m2}-E_{m1}$ and fitted by $E_{\mathrm{th2}}-E_{\mathrm{th1}}$.
(f) Angular dependence of the Zeeman-split spectral features at $h=0.7$. The split superconducting coherence peaks are denoted by $E_{z1}$ and $E_{z2}$. The splitting magnitude is $E_{z2}-E_{z1}$ and is fitted by $2\mathrm{H_{\parallel}}/\Delta_0$.
}
\label{DOS_Rashba}
\end{figure*}

To further clarify how momentum integration affects the spectral features, we explicitly calculate superconductors with Rashba and Ising SOC. For a superconductor with both SOC, its texture is $\mathbf{g}(\mathbf{p})=(-\alpha p_y, \alpha p_x, v\beta_{SO})$, where $\alpha$ and $\beta_{SO}$ are the Rashba and Ising SOC strength, and $v=\pm 1$ labels the two valleys \cite{xiao2012coupled,yuan2014possible}. We assume a circular Fermi surface with the Fermi energy much larger than all other relevant energy scales. The Rashba energy scale can then be approximated by $\alpha p_F$, where $p_F$ is the Fermi momentum. Taking $|\mathbf{p}|=p_F$, the momentum dependence reduces to the angular coordinate $\theta$, defined as the angle between $\mathbf{p}$ and the $\mathbf{x}$ axis. To avoid additional orbital effects associated with out-of-plane fields, we consider an in-plane magnetic field. Owing to the in-plane rotational symmetry, the field direction is chosen along the $x$ axis without loss of generality. For convenience, we also introduce following dimensionless parameters:
\begin{equation}
    h=\frac{H}{\Delta_0},\ t=\frac{k_BT}{\Delta_0},\ \alpha_{so}=\frac{\alpha p_F}{\Delta_0},\ \beta_{so}=\frac{\beta_{SO}}{\Delta_0},
\end{equation}
\noindent where $\Delta_0$ is superconducting gap under zero temperature and vanishing field. The physical gap is obtained self-consistently by minimizing the free energy \cite{altland2010condensed,wang2025unified,wang2025temperature,xie2020spin}, as detailed in the Supplemental Material \cite{supp}.

We first examine the field-dependent spectral features of a pure Rashba superconductor. For clarity, we consider a relatively strong Rashba SOC with $\alpha_{so}=3$. In the DOS calculation, a small intrinsic broadening $\eta=0.01\Delta_0$ is used throughout. To resolve these angular contributions, Figs.~\ref{DOS_Rashba}(a)-\ref{DOS_Rashba}(c) show the field evolution of the DOS at three representative momentum angles. The spin and magnetic-field directions are indicated by black and red arrows, respectively.

For $\mathbf{p}\parallel \mathbf{x}$ ($\theta=0$), the magnetic field is perpendicular to the SOC-pinned spin direction. At high fields, the DOS exhibits pronounced mirage gaps without Zeeman-split features. For $\mathbf{p}\parallel \mathbf{y}$ ($\theta=\pi/2$), the field is parallel to the spin direction, leading to the absence of a mirage gap and the emergence of Zeeman-split features. At intermediate angles ($0<\theta<\pi/2$ or $\pi/2<\theta<\pi$), the DOS exhibits both mirage gaps and Zeeman splitting under a magnetic field, as shown in Fig.~\ref{DOS_Rashba}(c). These angle-dependent behaviors are consistent with the general discussion above.

The overall DOS is obtained by integrating the angle-resolved DOS over all momentum angles [Fig.~\ref{DOS_Rashba}(d)]. Asymmetric V-shaped mirage gaps emerge outside the superconducting gap, in contrast to the symmetric U-shaped mirage gaps of Ising superconductors \cite{tang2021magnetic}. Moreover, with increasing magnetic field, the U-shaped superconducting gap is gradually transformed into a broadened V-shaped gap, although weak U-shaped features persist near the Fermi level. These V-shaped features arise from angular integration, as supported by the angular dependence of the mirage gaps and Zeeman-split features. 

\begin{figure*}[ht]
\centering
\includegraphics[width=0.95\linewidth]{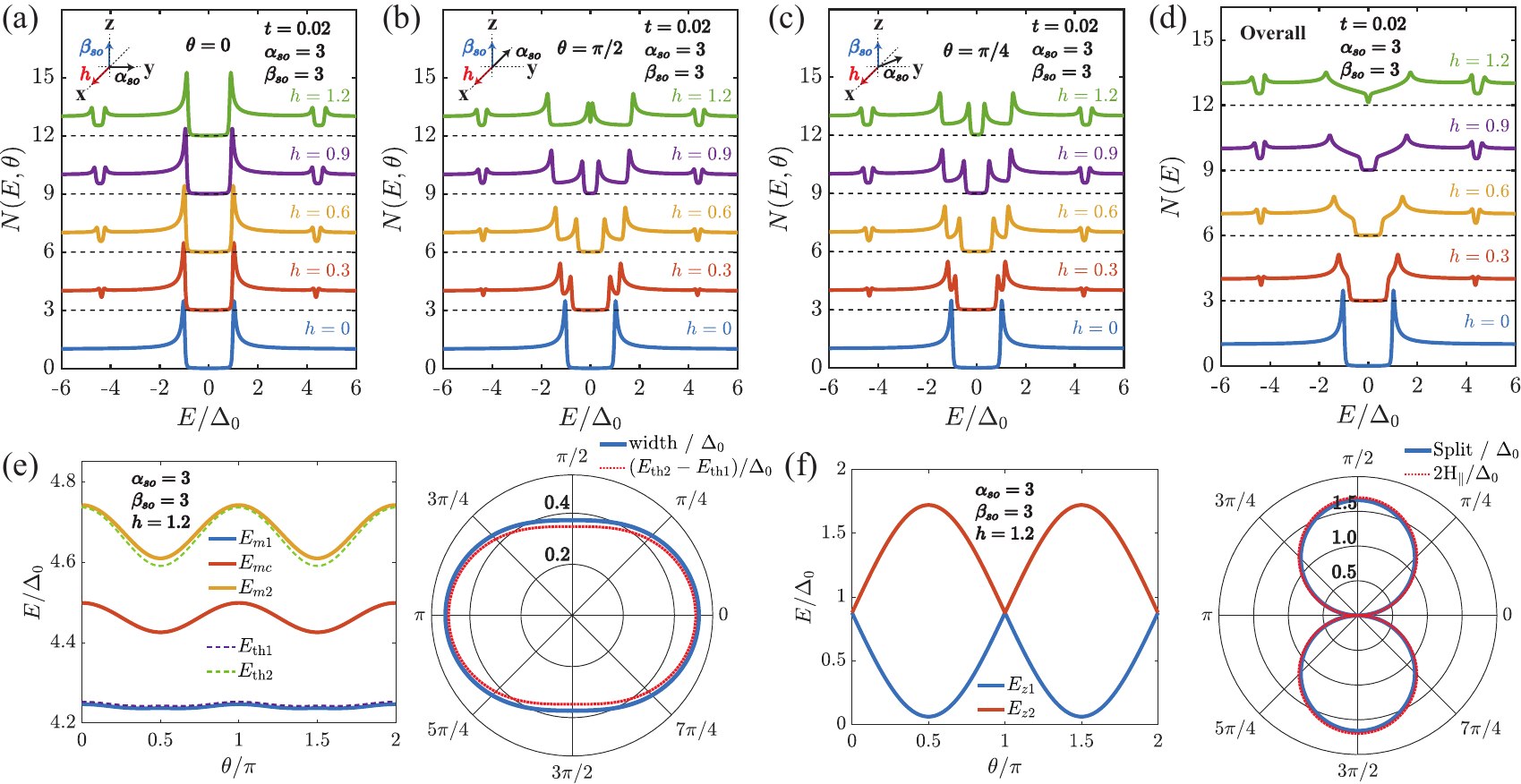}
\caption{
Magnetic-field-dependent spectral features of a superconductor with coexisting Rashba and Ising SOC with $\alpha_{so}=3$ and $\beta_{so}=3$. Field-dependent momentum-resolved DOS $N(E,\mathbf{p})$ at (a) $\theta=0$, (b) $\theta=\pi/2$, and (c) $\theta=\pi/4$. Blue arrows indicate the spin components pinned by Ising SOC. All other notations are the same as in Figs.~\ref{DOS_Rashba}(a)-\ref{DOS_Rashba}(c).
(d) Overall field-dependent DOS $N(E)$.
(e) Angular dependence of the mirage gap at $h=1.2$. The fitted coherence-peak positions are $E_{\mathrm{th1}}=\sqrt{(\alpha p_F)^2+\beta_{SO}^2+(\mathrm{H_\perp}-\Delta)^2}/\Delta_0$ and $E_{\mathrm{th2}}=\sqrt{(\alpha p_F)^2+\beta_{SO}^2+(\mathrm{H_\perp}+\Delta)^2}/\Delta_0$.
(f) Angular dependence of the Zeeman-split spectral features at $h=1.2$.
}
\label{DOS_Ising_Rashba}
\end{figure*}

Figures~\ref{DOS_Rashba}(e) and \ref{DOS_Rashba}(f) show the angular dependence of the mirage-gap positions and width at $h=0.7$, together with the positions of the two Zeeman-split peaks. Consistent with Figs.~\ref{DOS_Rashba}(a)-\ref{DOS_Rashba}(c), the mirage-gap width is maximal and Zeeman splitting is absent at $\theta=0$ and $\theta=\pi$, where the spin direction is perpendicular to the applied field. Conversely, at $\theta=\pi/2$ and $\theta=3\pi/2$, the mirage gap vanishes and the Zeeman splitting is maximal. The analytical results from Eqs.~(\ref{H_perp}) and (\ref{H_para}) accurately describe the mirage-gap [Fig.~\ref{DOS_Rashba}(e)] and Zeeman-split features [Fig.~\ref{DOS_Rashba}(f)], indicating that the mirage gap is mainly determined by the perpendicular component of the magnetic field, whereas the Zeeman splitting is governed by the parallel component. The small deviations may arise from the finite broadening $\eta=0.01\Delta_0$. Because both the mirage-gap width and Zeeman splitting vary continuously from zero to finite values with angle, angular integration naturally produces V-shaped spectral features. The asymmetry of the V-shaped mirage gap originates from the angular variation of its center position.

We further investigate the field-dependent spectral features of superconductors with coexisting Ising and Rashba SOC. To highlight their interplay, we consider equal strengths with $\alpha_{so}=\beta_{so}=3$. The angle-resolved DOS is shown in Figs.~\ref{DOS_Ising_Rashba}(a)-\ref{DOS_Ising_Rashba}(c). Unlike in the pure Rashba case, clear mirage gaps appear at every momentum angle upon applying an in-plane magnetic field. In addition, Zeeman-split spectral features emerge whenever the Rashba SOC has a component parallel to the applied field ($\theta\neq 0$ and $\theta\neq \pi$).

Near the Fermi level, the overall DOS [Fig.~\ref{DOS_Ising_Rashba}(d)] resembles that of the pure Rashba superconductor [Fig.~\ref{DOS_Rashba}(d)], whereas the shapes of their mirage gaps differ. These differences can be understood from the angular dependence of the mirage gap and Zeeman splitting, as shown in Figs.~\ref{DOS_Ising_Rashba}(e) and \ref{DOS_Ising_Rashba}(f). Because the Ising SOC is always perpendicular to the applied field, the mirage-gap width remains finite at all angles, with additional angle-dependent variations arising from Rashba SOC. Since these variations are relatively weak compared with the finite background width, angular integration produces a predominantly U-shaped mirage gap with only weak asymmetric V-shaped features, closely resembling the spectrum of an Ising superconductor. By contrast, the Zeeman-like splitting still varies from zero to a finite value with angle, producing an overall low-energy DOS similar to that of a pure Rashba superconductor. Notably, the angular dependence of both the mirage gap and Zeeman-split features are also consistent with the analytical results from Eqs.~(\ref{H_perp}) and (\ref{H_para}), further supporting the decomposition of the spectral response into perpendicular and parallel magnetic-field components.

\begin{figure}[ht]
\centering
\includegraphics[width=\linewidth]{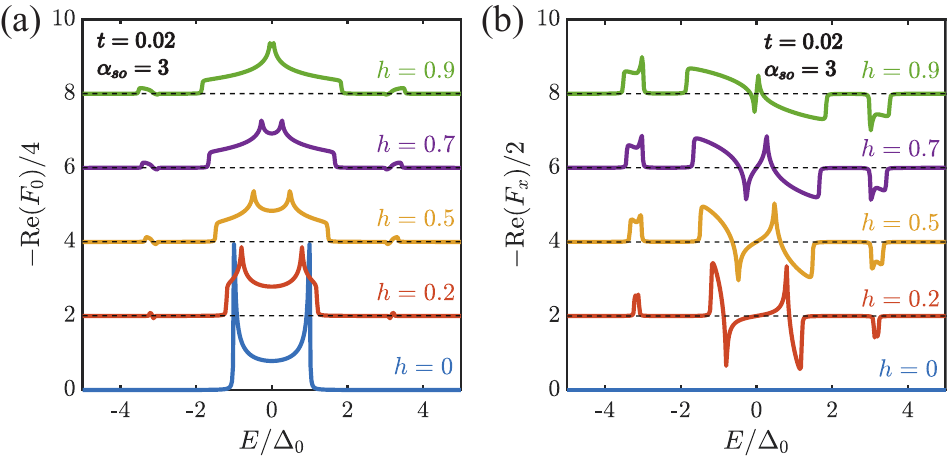}
\caption{
Field-dependent overall anomalous Green's function components of a pure Rashba superconductor with $\alpha_{\mathrm{so}}=3$. Field dependence of (a) -$\mathrm{Re}(F_0)/4$ and (b) -$\mathrm{Re}(F_x)/2$. Successive curves are vertically offset by 2, and their amplitudes are scaled for clarity.
}
\label{pairing_correlation}
\end{figure}

Finally, we examine the pairing correlations associated with the  overall mirage gaps through the anomalous Green's function $F(E)$.
Figure~\ref{pairing_correlation} presents the results for a pure Rashba superconductor, while other cases are discussed in the Supplemental Material \cite{supp}. Although all three triplet components $F_i$ are nonzero in the angle-resolved anomalous Green's function, only $F_0$ and $F_x$ remain nonzero after angular integration \cite{supp}. The nonzero anomalous Green's function components inside the mirage gaps demonstrate that these spectral features originate from finite-energy pairing correlations containing both spin-singlet and spin-triplet components, consistent with the mirage gaps in Ising superconductors \cite{tang2021magnetic}.

In summary, we have shown that mirage gaps can emerge in superconductors with general TRS SOC $\mathbf{g(p)}$, provided that the applied magnetic field has a component perpendicular to the SOC field. By contrast, a parallel component produces Zeeman-split spectral features. Notably, the mirage gap emerges at the characteristic energy scale $\sqrt{|\mathbf{g(p)}|^2+\mathrm{H_{\perp}^2}}$, which is distinct from the Zeeman-split scale near $\Delta \pm \mathrm{H_\parallel}$. These universal and field-dependent signatures not only enrich the understanding of finite-energy pairing \cite{tang2021magnetic,chakraborty2022interplay,bahari2022intrinsic,russmann2023interorbital,kornich2024emergence,bahari2024helical,wei2024gapless}, but also establish superconducting spectroscopy as a novel probe of SOC textures and strengths.

We have further investigated the spectral properties of superconductors with coexisting Rashba and Ising SOC under in-plane magnetic fields. We find that Rashba SOC produces asymmetric V-shaped mirage gaps associated with finite-energy pairing, whereas Ising SOC generates symmetric U-shaped mirage gaps. These features can be directly probed by scanning tunneling spectroscopy, providing a powerful route for determining the Rashba and Ising SOC components. Transition-metal dichalcogenide monolayers, such as $\mathrm{MoS_2}$ \cite{lu2015evidence,saito2016superconductivity} and $\mathrm{NbSe_2}$ \cite{xi2016ising,sohn2018unusual,yi2022crossover}, as well as ultrathin Pb films \cite{liu2018interface,wang2021ising}, provide promising platforms for testing our predictions.

\textit{Acknowledgments.--} X. Wang acknowledges stimulating discussions with K. T. Law, N. F. Yuan, and L. He. This work was supported by the Quantum Science and Technology National Science and Technology Major Project (Grant No. 2023ZD0300500), and the National Natural Science Foundation of China (Grant Nos. 52388201 and 12374048).



\nocite{*}

\bibliographystyle{Zou}
\bibliography{bib_main}

\end{document}